# CMOS compatible high-performance nanolasing based on perovskite-SiN hybrid integration

*Zhe He, Bo Chen ,Yan Hua, Zhuojun Liu, Yuming Wei, Shunfa Liu，An Hu，Xinyu Shen, Yu Zhang\*, Yunan Gao\*, Jin Liu\**

Z. He, B. Chen, Y. Hua, Z. Liu, Dr. Y. Wei, S. Liu, Prof. J. Liu
State Key Laboratory of Optoelectronic Materials and Technologies
School of Physics
Sun Yat-Sen University
Guangzhou, 510275, China
Email: liujin23@mail.sysu.edu.cn
Academic Website: http://spe.sysu.edu.cn/photon/

A. Hu, Prof. Y. Gao
State Key Laboratory for Mesoscopic Physics
School of Physics
Peking University
Beijing, 100871, China
Email: gyn@pku.edu.cn

X. Shen, Prof. Y. Zhang
State Key Laboratory of Integrated Optoelectronics
College of Electronic Science and Engineering
Jilin University
Changchun, 130012, China
Email: yuzhang@jlu.edu.cn

Abstract: Coherent light sources in silicon photonics are the long-sought holy grail because silicon-based materials have indirect bandgap. Traditional strategies for realizing such sources, e.g., heterogeneous photonic integration, strain engineering and nonlinear process, are technologically demanding. Here, we demonstrate a hybrid lasing device composing of perovskite nanocrystals and silicon nitride nanobeam cavity. We fabricate SiN photonic crystal naonobeam cavities on a solid substrate with significantly improved thermal and mechanical stabilities compared to conventional suspended ones. In addition, adding a PMMA-encapsulation layer on top of the SiN can significantly boost the Q-factor of the cavity mode. By dispersing perovskite nanocrystals as emitters in the PMMA layer, we obtained high-performance coherent emissions in terms of lasing threshold, linewidth and mode volumes. Our work offers a compelling way of creating solution-processed active integrated photonic devices based on the mature platform of silicon photonics for applications in optical information science and photonic quantum technology.

## 1. Introduction

Silicon Photonics has been long recognized as a very powerful platform for future broadband and high-speed data transmission and processing, because of its excellent scalability and compatibility with modern complementary metal oxide semiconductor (CMOS) technology.[1-3] The success of silicon photonics has fueled rapid developments in new research directions ranging from refractive index to spectroscopic sensing,[4] which requests the extension of the operation wavelength from silicon materials' transparent window of 1.1 μm to 4 μm to visible and near-infrared range. Therefore, silicon nitride (SiN) with wide transparency window, from visible to near-infrared, has been included in silicon photonics as a complementary integrated photonic platform.[5] Despite impressive developments in the core components, such as extremely low-loss waveguide,[6] ultrafast modulators[7] and broadband detectors,[8] the coherent light sources that can be easily implemented in the standard fabrication process are still highly desirable and yet illusive.[9, 10]

Last two decades have witnessed tremendous efforts in realizing coherent light sources in silicon photonics. Hybrid photonic integration by either wafer bonding[11] or growing III-V material on silicon has enabled electrically pumped lasers,[12, 13] however this approach is rather costly and complicated due to lattice mismatch between silicon and III-V materials. By engineering strains, germanium (Ge), compatible with CMOS process, can be tuned from the indirect bandgap to the direct bandgap for achieving lasing emission, yet such devices are mostly working at cryogenic temperature with a limited radiative efficiency.[14] Recently, exploring nonlinear optical processes is emerging as an elegant solution, e.g., using stimulated Raman emission[15] or frequency comb generation[16]. The difficulty in this direction lies in the electrical injection and efforts on indirect electrically pumped comb are being heavily pursued.[17]

Alternatively, solution-processed gain materials could be introduced to silicon based photonic nanostructures to achieve light emitting devices that could be easily implemented with the standard fabrication process.[18-24] As a direct-bandgap semiconductor, colloidal all-inorganic perovskite nanocrystals are a compelling gain material for laser applications, whose emission wavelength can cover the whole visible range by changing the composition stoichiometry while maintaining high photoluminescence quantum yield.[25, 26] So far, lasing actions have been mainly achieved by using micro-cavities made from solution-processed thin films in which the perovskite material serves as both the lasing cavity and the gain material. In such cases, laser cavities are strictly restricted to several types of particular cavity geometries, and the optical gain suffers from the low quantum efficiency due the polycrystalline nature of the perovskite thin films.[18, 24, 27] Combining perovskite nanocrystals and silicon based photonic cavities may enable lasing actions with improved performances in terms of device footprint, lasing threshold and the emission linewidth, which could serve as a viable alternative for coherent light sources in silicon photonics.[28]

In this work, we demonstrate on-substrate photonic crystal nanobeam cavities with improved mechanical and thermal stabilities over the widely used suspended nanobeam cavities. Our nanobeam cavities exhibit increased Q-factors when coated with a layer of polymer, offering a compelling way of integrating solution-processed emitters. Efficient coupling between perovskite nanocrystals and the nanobeam cavity is characterized by time-resolved photoluminescence measurements. The spontaneous emission rate of perovskite nanocrystals is significantly enhanced when their emission couples to the cavity mode. We've achieved a highly compact device with the state of the art lasing threshold (5.62 µJ/cm$^2$) and an ultra-narrow linewidth (0.045 nm).

## 2. Design and Characteristics of the SiN nanobeam

**Figure 1**a depicts the structure of the proposed device in which a SiN nanobeam is embedded between a silicon oxide substrate and a top layer of poly (methyl methacrylate) (PMMA) hosting perovskite nanocrystals. The absorption and photoluminescence (PL) spectra of the CsPb(Br/I)$_3$ perovskite nanocrystals dispersed in toluene solution are presented in **Figure 1**b, showing PL peak at ~ 680 nm. **Figure**

**1**c shows a transmission electron microscopy (TEM) image of the perovskite nanocrystals. The perovskite nanocrystals have an average size of 20 nm with a cubic shape.

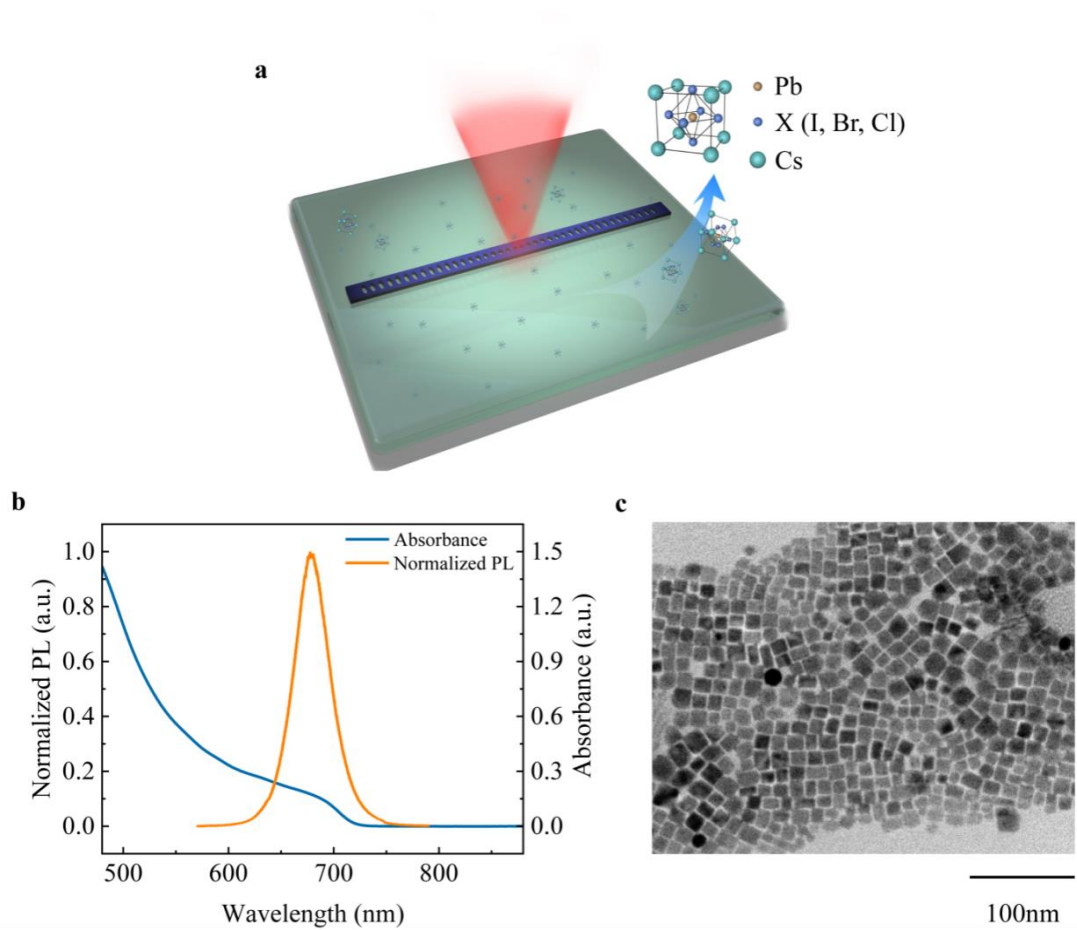

**Figure 1.** Proposed nanobeam laser devices and characterizations of perovskite nanocrystals. (a) Schematic of the lasing devices consisting of a SiN nanobeam on a silicon oxide substrate covered by a layer of PMMA mixed with perovskite nanocrystal. (b) Absorption and photoluminescence (PL) spectrum of the perovskite nanocrystals. (c) Transmission electron microscopy image (TEM) of perovskite nanocrystals.

We optimized parameters of the on-substrate nanobeam photonic crystal cavities[29, 30] via finite difference time domain (FDTD) calculations. As shown in **Figure 2**a, the cavity consists of a tapered section in the middle and two Bragg sections on both sides, supported on a silicon oxide substrate. Such on-substrate nanobeams offer greatly improved mechanical and thermal stabilities over the conventional suspended ones. In addition, after coated with a layer of PMMA with the same refractive index as the silicon oxide, the Q-factor of the cavity mode can be significantly increased, facilitating the onset of the lasing operation. **Figure 2**b shows the electric field ($E_y$) distribution of the fundamental mode of the nanobeam cavity in the *xy* plane. The calculated cavity Q-factor is about 200,000 with resonance centered at 684 nm (Figure S1, Supporting Information).

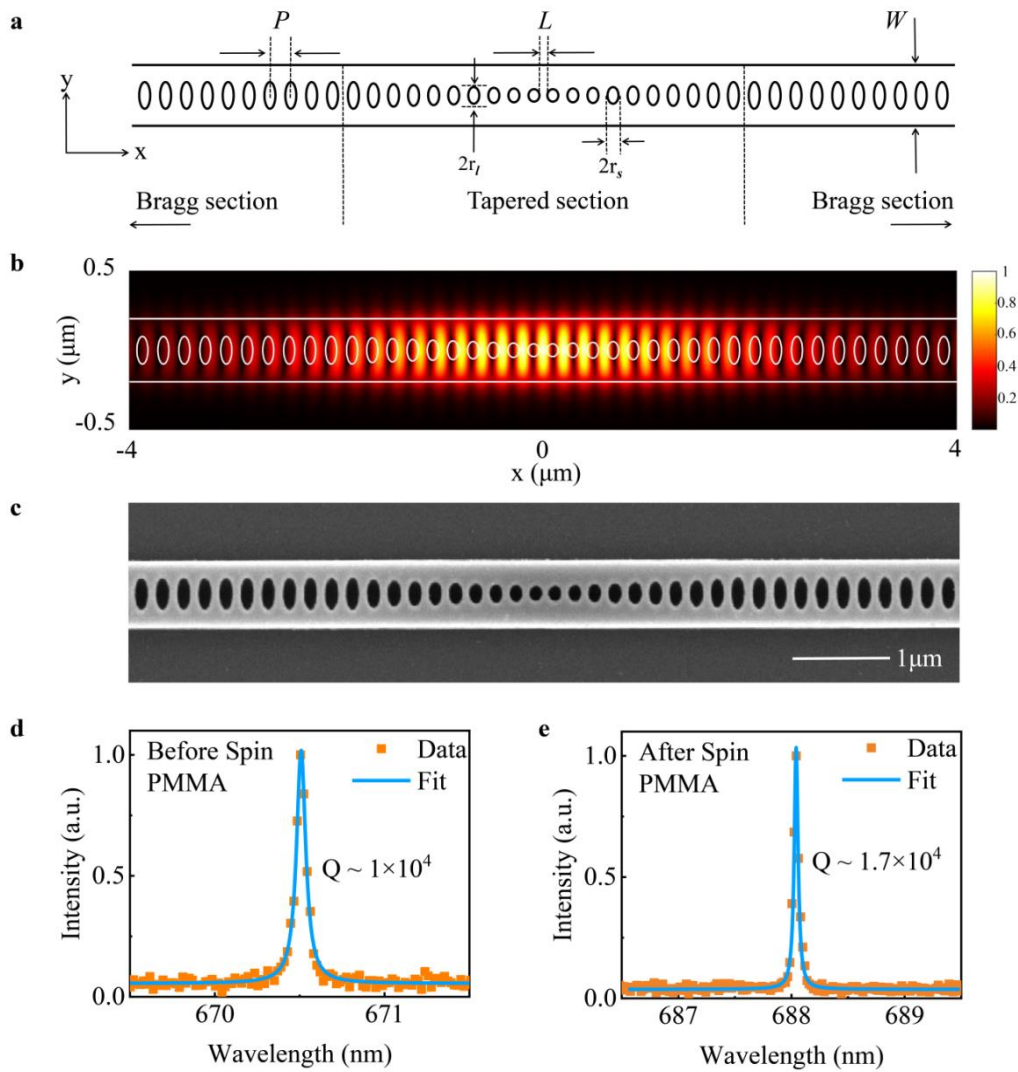

**Figure 2.** Design and Characterizations of the SiN nanobeam. (a) Schematic of the nanobeam device with elliptical holes, which consists of one linearly tapered central region and two constant Bragg sections. (b) Electric field $|E_y|^2$ component in the *xy*

plane of the fundamental cavity mode. (c) Top view scanning electron microscope (SEM) image of the nanobeam structure. (d) Measured PL spectrum of the cavity mode before spin-coating PMMA (Q ~ $1.0 \times 10^4$) and (e) after spin-coating PMMA (Q ~ $1.7 \times 10^4$). The results indicate that Q-factor of the cavity mode can be significantly boosted with an organic polymer coating.

The nanobeam cavities were fabricated in a 240-nm-thick SiN membrane supported on a 2 μm silicon dioxide layer, and details of the fabrication are described in Methods. The scanning electron microscopy (SEM) image of a fabricated SiN cavity on thermal oxide is shown in **Figure 2**c. Since the refractive index (n ~ 1.47) of PMMA is roughly the same as the silicon oxide substrate, we use it as the coating layer above the cavity to create a symmetric refractive index distribution along the z direction. We measured the cavity Q-factors for the nanobeam before and after PMMA encapsulation by utilizing the cavity enhanced intrinsic fluorescence of SiN[31] without the introductions of grating couplers and adiabatic tapers. The avoidance of the grating couplers and adiabatic tapers allows the fabrication of more Bragg mirrors, which boosts the achievable Q-factor more than one order of magnitude respective to the previous work[30] . **Figure 2**d presents the PL spectrum of the cavity mode before the coating of the PMMA layer, showing a sharp resonance at 670.5 nm with the Q-factor of 10,000. The discrepancy between the simulated and measured Q-factors is attributed to the imperfections from the fabrication, including nonuniformity of the elliptical hole sizes and side-wall roughness, etc. For cavities with PMMA encapsulation, the resonant wavelength red-shifted to 688 nm, and the Q-factor increased to 17,000 **(Figure 2**e**)**. These phenomena can be well reproduced in the FDTD simulations as shown in Figure S1. We simulated a series of refractive index of the top encapsulating layer, and the simulated result indicates that a symmetric refractive index distribution along the z direction gives the highest Q-factor of the nanobeam (Figure S2, Supporting Information). In addition, we systematically varied the cavity's geometries, by linearly scaling the hole radius and periods, and found the cavity modes can be tuned continuously across the entire visible range (Figure S3, Supporting Information).

### 3. Coupling perovskite nanocrystals to PMMA-encapsulated SiN nanobeam

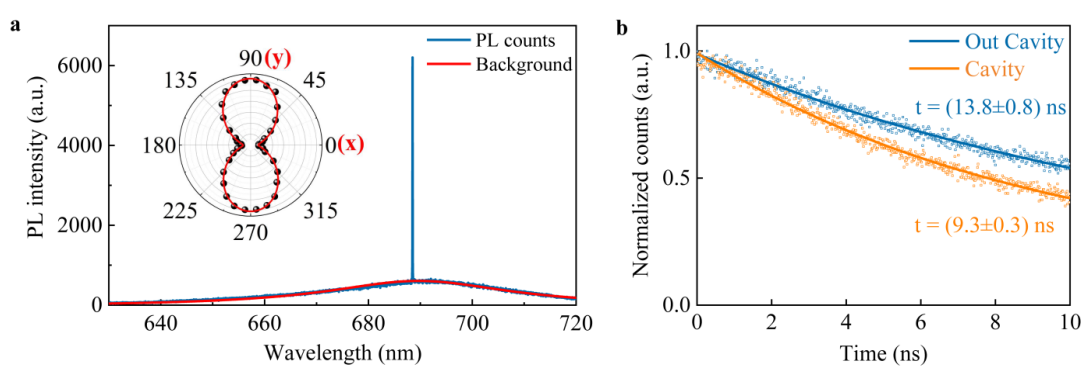

**Figure 3.** The coupling between the perovskite nanocrystals and the cavity mode of the nanobeam. (a) Emission spectra from perovskite nanocrystals on the cavity (blue) as compared to the perovskite PL without a cavity (red). Spectra are measured at an energy density of 20.0 µJ/cm$^2$. Inset: polarization characteristic of the fundamental mode of the nanobeam. The same coordinate system as in Figure 2 is used. (b) Normalized time-resolved PL from perovskite nanocrystals on the cavity (yellow) and on an unpatterned region (blue). The dots are the raw experimental data. An averaged Purcell factor of 1.48 is extracted.

$CsPb(Br/I)_3$ perovskite nanocrystals were employed as the optical gain medium for the realization of the coherent light sources. The nanocrystal solution with PMMA added was spin-coated on the nanobeam samples. While pure perovskite nanocrystals have a refractive index of ~ 2.6, the thin film has a refractive index close to PMMA due to the a low volume fraction of the nanocrystals. Efficient coupling between the perovskite nanocrystals and the SiN nanobeam cavity was observed, as shown in the

**Figure 3**a which a sharp cavity resonance in the PL spectrum was observed only in the cavity regime (blue line). Such a sharp peak was absent in the PL spectrum measured from the area without the cavity (red line). The observed sharp emission peak was strongly polarized in y-direction (perpendicular to the nanobeam) as shown in the inset of Figure 3a, and is consistent to the simulated polarization of the cavity mode. To quantitatively characterize the coupling between the perovskite nanocrystals and the nanobeam cavity, we measured the lifetime of the spontaneous emission from the perovskite nanocrystals at a low pumping intensity of 0.2 µJ/cm². **Figure 3**b shows the PL decay traces of the perovskite nanocrystals in and out of the cavity regime. Under the low excitation power, only exciton states are prepared and their decay follows a single-exponential function:

$$I(t) = I_0 + Ae^{-(t/\gamma_0)}$$

where $I(t)$ is the PL intensity at time $t$, $I_0$ is the background intensity, $\gamma_0$ is the decay constant of excitons in perovskites and $A$ is a scaling constant.

As shown in **Figure 3**b, the lifetime for the uncoupled case is 13.8 ± 0.8 ns, and for the coupled case is 9.3 ± 0.3 ns. A Purcell factor is calculated of 1.48 ± 0.13, indicating enhancement of the spontaneous emission. To compare the measured Purcell enhancement with the theoretical calculation based on FDTD simulations, we employed the following equation for calculating the Purcell factor:

$$F_{max} = 1 + \frac{3\lambda^3}{4\pi^2 n^2} \frac{Q_p}{V} \psi(r)$$

The expression represents the Purcell factor of the emitter located on the nanobeam surface at position of maximum field intensity, where $\lambda$ is the resonance wavelength, $n$ is the refractive index of SiN at the cavity resonance, $Q_p$ is the Q-factor of the perovskite emission linewidth, $V$ is the cavity mode volume, and $\psi(r)$ is the ratio of the mode intensity at the emitter's location over the maximum. We note that we used the Q-factor of the emitter instead of the cavity since our system is in the bad emitter regime, where the linewidth of the emitter is much broader than the cavity linewidth. In our cases, the linewidth of the perovskite nanocrystal emission is 30 nm, giving rise to a Q-factor about 23 (Figure 1b). From the FDTD simulations, we obtain $V \approx 2.5 \left(\frac{\lambda}{n}\right)^3$ and $\psi(r)$ is 0.39, assuming that the emitter is located at the center of the nanobeam cavity surface. With these values, the resulting theoretically calculated Purcell factor is 1.54, consistent with the measured number in the experiment.

## 4. Optical characterization of lasing

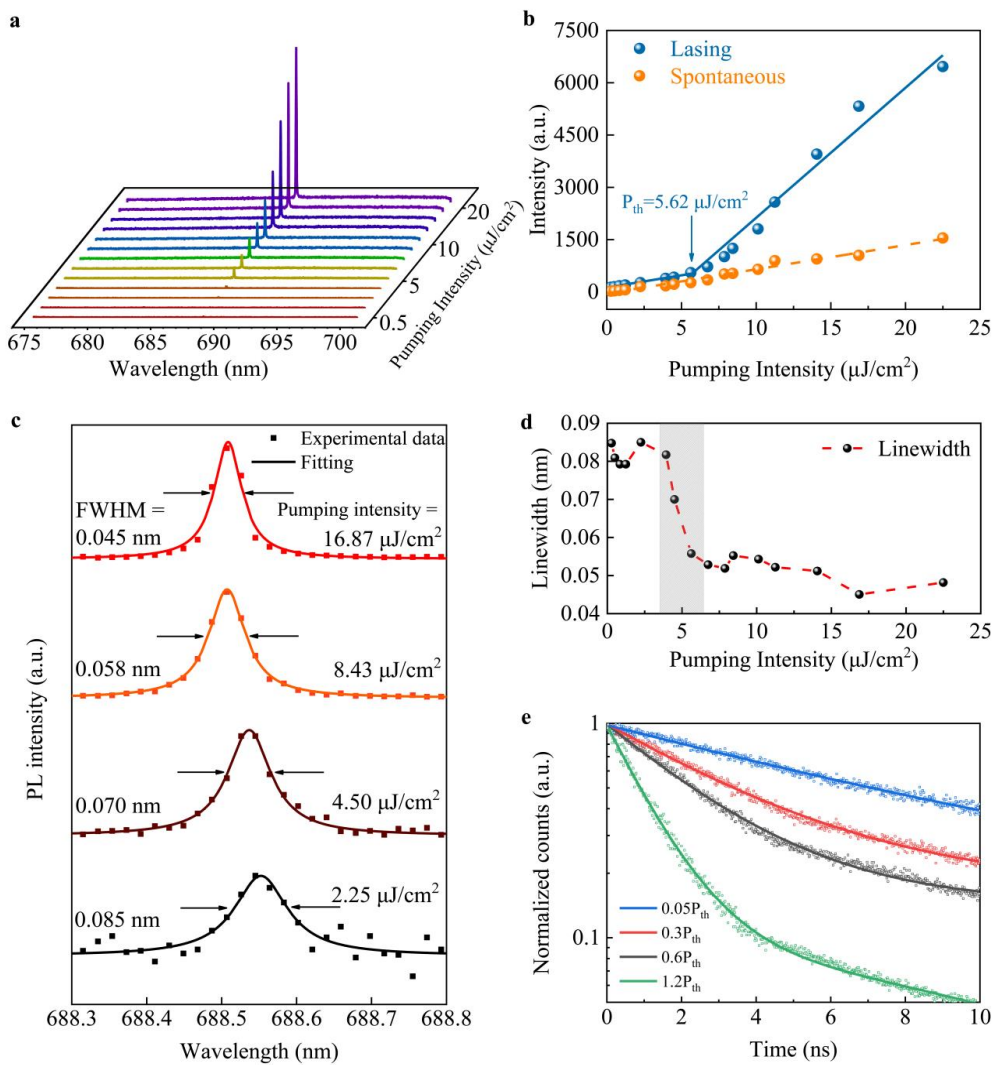

Figure 4. Characterizations of the perovskite nanocavity lasing. (a) PL spectra of the devices at different pumping intensities. (b) L-L curve, the cavity emission shows a nonlinear increase of the output intensity, while the background emission shows a linear dependence on the pump intensity. (c) Lorentzian fittings of the cavity modes at different pumping intensities (d) Linewidths of the cavity modes with the increase of the pumping intensity. (e) Pump-intensity-dependent PL decay curves of the device.

To identify the transition from spontaneous emission to lasing, we investigated the characteristics of the cavity mode at different pump intensities. Pump intensity dependent spectra are presented in **Figure 4**a. With the increase of the pump intensity, a clear lasing peak at 688 nm emerges and the cavity mode became dominating in the spectra compared to the background emission. **Figure 4**b shows the intensity change of the cavity mode as a function of pump intensity, referred to as the "light input–light output" or "L-L" curve, in which a clear tuning point can be identified. The spontaneous emission from an "off-resonant" position at the same wavelength, which does not show lasing, is also plotted in Figure 4b for comparison. The fitting of "L-L" curve with a rate equation model in logarithm scale is presented in Figure S4, resulting in a β-factor of 0.08. **Figure 4**(c,d) present the dependence of linewidth of the cavity mode on the pumping intensity. The spectra of the cavity modes at different pump intensities are fitted by Lorentzian line-shapes. With the increase of the pump intensity from 2.25 to 16.87 μJ/cm$^2$, the linewidth of the cavity mode narrowed down from 0.085 to 0.045 nm. We notice that the cavity mode slightly blue-shifted at the high pump intensity, which can be attributed to the change of the refractive index of the nanobeam due to the increased carrier density[32] and the heating of PMMA.[33] The relatively small blueshift of the cavity mode above threshold is a good indication of the improved thermal stability due to the on-substrate structure. The linewidth of the cavity mode as a function of the pump intensity is plotted in **Figure 4**d, showing a clear linewidth narrowing behavior. The rapid linewidth reduction regime, shaded in gray, synchronizes with the regime of dramatic change in the output intensity, which indicates the transition from spontaneous emission to stimulated emission. We note that the thermal optics induced linewidth broadening at high pumping intensities is not particularly obvious in our experiment, which could be due to two reasons. First, the thermal stability of our device is greatly improved by employing the on-substrate cavity. Second, with the low threshold, the energy of the laser pulses (above threshold) we injected into the device is relatively low, elevating significant thermal-optics effect. We note that the linewidth of the cavity mode above threshold is close to the resolution of our spectrometer. Our lasing devices exhibit superior performances with the state-of-the-art lasing threshold down to 5.62 μJ/cm$^2$, the lasing linewidth as narrow as 0.045 nm, small footprint of 4 μm$^2$ and the CMOS compatibility, as shown in **Table 1**.

**Table 1. Comparison of the performance for the existing all-inorganic perovskite nanolasing.**

| Reference | Nanostructure Type/Cavity Architecture | Lasing Wavelength (nm) | Lasing Threshold (μJ/cm$^2$) | Linewidth (nm) | Hot cavity Q-Value [a] | Device Size (CSA × Height) [b] | Scalable [c] | Pumping Source |
|---|---|---|---|---|---|---|---|---|
| [34] | QDs/WGM | 524.5 | 11000 | 0.32 | ~1600 | ≥ 1960 um$^2$ × NA [d] | × | 100 fs/400 nm |
| [35] | QDs/DBR | 504 | 11 | 0.6 | ~840 | ≥ 6300 um$^2$ × 4 um [e] | × | 100 fs/400 nm |
| [36] | Thin film/WGM | 427 | 7.2 | 0.8 | ~530 | ≥ 6.25 um$^2$ × 0.6um | × | 150 fs/400 nm |
| [37] | Micro-cubes/FP | 577 | 439 | 0.46 | ~1150 | ≥ 0.25 um$^2$ × 0.5 um | × | 35 fs/800 nm |
| [38] | QDs/DBR | 522 | 0.39 | 0.9 | ~580 | ≥ 20 cm$^2$ × 5 um | × | 50 fs/400 nm |
| [39] | Nanoplatelets/DBR | 427.5 | 12 | 1.43 | ~300 | ≥ 490 um$^2$ × 2.4um [f] | × | 100 fs/375 nm |
| [40] | Nanocuboids/FP | 539 | 40.2 | 0.29 | ~2075 | ≥ 0.16 um$^2$ × 0.4um | × | 100 fs/400 nm |
| [41] | Microsheets/FP | 542.7 | 17 | 0.41 | ~1300 | ≥ 80 um$^2$ × 5 um | × | 100 fs/400 nm |
| [42] | Thin film/DFB | 653.7 | 33 | 4.9 | ~130 | ≥ 100 mm$^2$ × 0.16 um | × | 90 ps/355 nm |
| [43] | NWs/FP | 538 | 6.2 | 0.26 | ~2069 | ≥ 100 um$^2$ × 1 um | × | 150 fs/402 nm |
| [44] | NWs/FP | 535 | 4 | 0.2 | ~2675 | ≥ 0.28 um$^2$ × 10 um | × | 100 fs/400 nm |
| [45] | Microcubic/FP | 536.5 | 203.6 | 0.053 | ~10100 | ≥ 2.89 um$^2$ × 1.7 um | × | 40 fs/800 nm |
| This work | NCs/PhC | 688.5 | 5.62 | 0.045 | ~15000 | ≥ 4 um$^2$ × 0.24 um | √ | 5 ps/532 nm |

a) The "hot cavity" Q-factor is derived from the lasing mode. b) We calculate the device size from its cross-sectional area (CSA) and height of the devices. c) scalable by using CMOS compatible process. d) NA denotes for the Non-Available Information. e,f) Since the CSA of the these DBRs are not given directly, we use the spot area as the CSA to calculate the size. The abbreviations are as follows: WGM: The Whispering Gallery Mode Cavity; DBR: Distributed Bragg reflector; NWs: Nanowires; DFB: Distributed feedback; FP: Fabry-Perot cavity; NCs: Nanocrystals; PhC: Photonic crystal.

In addition, we measured the lifetime of the cavity mode in relation to the pumping intensity **(Figure 4e)**. Under a pump intensity of much lower than the threshold (0.05 $P_{th}$), the PL decay trace followed a slow single-exponential decay, resembling the spontaneous emission trace (blue trace). With the increase of pump intensity approaching the threshold ($0.3P_{th}$, $0.6P_{th}$), the decay got faster and can be fitted by a biexponential decay function. The faster component is due to the Auger recombination because of high density of photon excited excitons,[46] see the extracted amplitude and lifeitme of the fast component as a function of the pumping intensity in Figure S5. As the pumping intensity surpassed the threshold (1.2 $P_{th}$), the PL decay time decreased to 1 ns (green trace). The significantly shortened PL lifetime above the threshold is a result of onset of stimulated emission.[47]

## 5. Discussion and conclusions

In conclusion, we have demonstrated the first high-performance perovskite nanocrystal lasing based on the SiN nanobeam cavity, which could serve as a potential coherent light sources to extend silicon photonics to the visible regime. This approach greatly benefits from the advantages of non-suspended structure and the symmetric distribution of the refractive index along z direction, offering substantial improvements for both mechanical and thermal stabilities and the Q-factor of the lasing cavity. Thanks to the high material gain from perovskite crystals, the high-Q and low mode volume of the nanobeam cavity, our highly compact lasing device simultaneously exhibit a ultra-low threshold of 5.62 μJ/cm$^2$ and a record narrow linewidth (0.045 nm), which is potentially applicable for high-resolution spectroscopy and coherent sensing. Unlike gain materials with fixed luminous wavelengths, such as two-dimensional materials, we can create laser devices spanning the entire visible range by controlling material composition and linearly scaling size of the photonic cavities. By reducing the density of the perovskite nanocrystals, our device can be explored to build coherent single-photon sources[48] with deterministic positioning techniques[30] and thermally tunable cavity[33], which may potentially change the landscape of solid-state quantum photonics with the replacements of epitaxial quantum dots[49] by low-cost chemically synthesized perovskite nanocrystals. Our work paves a credible way towards highly-efficient active integrated photonic devices based on silicon photonics platform, with applications in optical information science and quantum photonic technology.

## 6. Materials and Methods

*Perovskite nanocrystals preparation.*

Chemicals: caesium carbonate ($Cs_2CO_3$, Aldrich, 99.9%), 1-octadecene (ODE, Sigma-Aldrich, 90%), oleic acid (OA, Sigma-Aldrich, 90%), oleylamine (OAm, Acros Organics, 80%–90%), lead bromide ($PbBr_2$, macklin, 99.99% metals basis), lead iodide ($PbI_2$, ThermoFisher, 99.9985%), toluene (Fisher Scientific, HPLC grade).

Synthesis: For the preparation of Cs-oleate precursor, a mixture of $Cs_2CO_3$ (0.407 g), OA (1.25 mL) and ODE (15 mL) was loaded into a 50 mL 3-neck flask and dried under vacuum for 1 h at 120 °C. After $Cs_2CO_3$ was completely dissolved, the reaction temperature was raised up to 150 °C under nitrogen for 2 h. Then the Cs-oleate precursor cooled to room temperature naturally, and was kept in a glovebox. For the synthesis of $CsPb(Br/I)_3$ nanocrystals, 0.320 mmol $PbI_2$, 0.056 mmol $PbBr_2$ and ODE (10 mL) was loaded into a 50 mL 3-neck flask and dried under vacuum for 1 h at 120 °C. Followed by the injection of OA (1 mL) and OAm (1 mL) under nitrogen, the solution temperature was elevated to 170 ℃ and 0.8 ml of Cs-oleate, pre-heated at 100 °C, was quickly injected into this reaction mixture. After 5 s, the reaction flask was immediately cooled to room temperature with an ice-bath. Finally, the CsPb(Br/I)$_3$ nanocrystals were purified by centrifugation (5500 rpm, 8 min) and the precipitate was dispersed in toluene to form a stable solution.

*Fabrication of the lasing devices.* The nanobeam cavities were fabricated on a 240 nm thick SiN membrane grown via PECVD (Oxford PlasmaPro System100 ICP180-CVD) supported on a 2 μm silicon dioxide layer. A 400 nm ARP-6200 electron beam resist was spin-cast (4000 rpm, 60 s) on top of a 1 x 1 cm$^2$ SiN membrane, and baked at 150 ℃ for 3 minutes. The pattern was defined in electron beam resist by electron-beam lithography (Raith Vistec EBPG5000+ 100kV), and then transferred into the SiN membrane using reactive ion etching system (Oxford PlasmaPro System100RIE) with CHF$_3$/O$_2$ gas. The residual resist was removed by a gentle oxygen plasma RIE process.

In order to ensure the quality of the device, the proportion of solution is an important parameter to optimize. Too high concentration will obviously reduce the Q-factor of the cavity (also see Figure S2). Here, we injected a 10 ml toluene to 1ml CsPb(Br/I)3 solution and then mixed PMMA with a volume ratio of 1:2 as the gain medium we use. Lastly, the gain medium solution was spin-coated on the nanobeam sample at 3000 rpm for 1 minute, followed by a thermal annealing at 60 ℃ for 5 minutes.

*Optical measurement.* The fabricated sample was characterized on a home-built

PL setup sketched in the supporting information (Figure S6). For the cavity measurements, a continuous wave (CW) laser with a single mode fiber excitation at 532 nm was focused (Olympus 50 × objective lens, NA = 0.90) onto the sample center. The spectra were collected by a spectrometer with a 1200 line/mm grating (Princeton Instrument SP2758). The cavity emission signal was filtered out spectrally by a 532 nm long-pass filter. Since the low-efficiency of intrinsic fluorescence of SiN, a long exposure time (5-10 s) were employed. The same setup has been used to characterize nanobeam coated with perovskite, but with a ps laser (532 nm, 5 ps, 86 MHZ) instead of a CW laser. The polarization analysis was performed by a half-wave plate and a polaroid. For time-resolved decay measurements, a Ti-sapphire pulsed laser with a pulse duration of 120 fs and repetition rate of 79 MHz is used to excite the samples. The sample is excited by the optical pulses through an objective with NA = 0.65, and signal detected with an avalanche photon detector connected to a single photon counting module (PicoHarp 300). The scattered laser background is suppressed with the grating of spectrometer.

## Supporting Information

Supporting Information is available from the Wiley Online Library or from the author.

## Conflict of interest

The authors declare no conflict of interest.

**Acknowledgements**

Zhe He and Bo Chen contributed equally to this work.

This research was supported by National Key R&D Program of China (2018YFA0306100), the National Natural Science Foundation of China (11874437, 61935009), Guangzhou Science and Technology Project (201805010004), the Natural Science Foundation of Guangdong (2018B030311027, 2016A030306016, 2016TQ03X981), the Fundamental Research Funds for the Central Universities and the national super- computer center in Guangzhou.

**References**

[1] M. A. Taubenblatt, *J. Lightwave Technol*. **2011**, *30*, 448.

[2] B. Jalali, S. Fathpour, *J. Lightwave Technol*. **2006**, *24*, 4600.

[3] J. S. Levy, A. Gondarenko, M. A. Foster, A. Turner-Foster, A.L. Gaeta, M. Lipson, *Nat. Photonics* **2010**, *4*, 37.

[4] A. Z. Subramanian, E. Ryckeboer, A. Dhakal, F. Peyskens, A. Malik, B. Kuyken, H. Zhao, S. Pathak, A. Ruocco, A. D. Groote, P. Wuytens, D. Martens, F. Leo, W. Xie, U. D. Dave, M. Muneeb, P. V. Dorpe, J. V. Campenhout, W. Bogaerts, P. Bienstman, N. L. Thomas, D. V. Thourhout, Z. Hens, G. Roelkens, R. Baets, *Photonics Res*. **2015**, *3*, B47.

[5] A. Rahim, E. Ryckeboer, A. Z. Subramanian, S. Clemmen, B. Kuyken, A. Dhakal, A. Raza, A. Hermans, M. Muneeb, S. Dhoore, Y. Li, U. Dave, P. Bienstman, N. L. Thomas, G. Roelkens, D. V. Thourhout, P. Helin, S. Severi, X. Rottenberg, R. Baets, *J.*

*Lightwave Technol.* **2017**, *35*, 639.

[6]  J. F. Bauters, M. J. R. Heck, D. John, D. Dai, M. C. Tien, J. S. Barton, A. Leinse, R. G. Heideman, D. J. Blumenthal, J. E. Bowers, *Opt. Express* **2011**, *19*, 3163.

[7]  S. F. Preble, Q. Xu, B. S. Schmidt, M. Lipson, *Opt. Lett.* **2005**, *30*, 2891.

[8]  F. Schuster, D. Coquillat, H. Videlier, M. Sakowicz, F. Teppe, L. Dussopt, B. Giffard, T. Skotnicki, W. Knap, *Opt. Express* **2011**, *19*, 7827.

[9]  H. Rong, R. Jones, A. Liu, O. Cohen, D. Hak, A. Fang, M. Paniccia, *Nature* **2005**, *433*, 725.

[10] L. Pavesi, L. D. Negro, C. Mazzoleni, G. Franzò, F. Priolo, *Nature* **2000**, *408*, 440.

[11] J. V. Campenhout, P. Rojo-Romeo, P. Regreny, C. Seassal, D. Van Thourhout, S. Verstuyft, L. Di Cioccio, J.-M. Fedeli, C. Lagahe, R. Baets, *Opt. Express* **2007**, *15*, 6744.

[12] Z. Wang, B. Tian, M. Pantouvaki, W. Guo, P. Absil, J.V. Campenhout, C. Merckling, D. V. Thourhout, *Nat. Photonics* **2015**, *9*, 837.

[13] D. Jung, J. Norman, M. J. Kennedy, C. Shang, B. Shin, Y. Wan, A. C. Gossard, J. E. Bowers, *Appl. Phys. Lett.* **2017**, *111*, 122107.

[14] H. Liu, T. Wang, Q. Jiang, R. Hogg, F. Tutu, F. Pozzi, A. Seeds, *Nat. Photonics* **2011**, *5*, 416.

[15] Y. Takahashi, Y. Inui, M. Chihara, T. Asano, R. Terawaki, S. Noda, *Nature* **2013**, *498*, 470.

[16] G. Kurczveil, D. Liang, M. Fiorentino, R. G. Beausoleil, *Opt. Express* **2016**, *24*,

16167.

[17] B. Stern, X. Ji, Y. Okawachi, A. L. Gaeta, M. Lipson, *Nature* **2018**, *562*, 401.

[18] H. Cha, S. Bae, M. Lee, H. Jeon, *Appl. Phys. Lett.* **2016**, *108*, 181104.

[19] Z. Yang, M. Pelton, I. Fedin, D. V. Talapin, E. Waks, *Nat. Commun*. **2017**, *8*, 1.

[20] S. Gupta, E. Waks, *Opt. Express* **2013**, *21*, 29612.

[21] Y. Wang, C. Yuan, Y. Yang, M. Wu, J. Tang, M. Shih, *Nano Rev.* **2011**, *2*, 7275.

[22] J. Yang, J. Heo, T. Zhu, J. Xu, J. Topolancik, F. Vollmer, R. Ilic, P. Bhattacharya, *Appl. Phys. Lett.* **2008**, *92*, 261110.

[23] C. Fong, Y. Yin, Y. Chen, D. Rosser, J. Xing, A. Majumdar, Q. Xiong, *Opt. Express* **2019**, *27*, 18673.

[24] Z. Li, J. Moon, A. Gharajeh, R. Haroldson, R. Hawkins, W. Hu, A. Zakhidov, Q. Gu, *ACS Nano*, **2018**, *12*, 10968.

[25] L. Protesescu, S. Yakunin, M. I. Bodnarchuk, F. Krieg, R. Caputo, C. H. Hendon, R. Yang, A. Walsh, M. V. Kovalenko, *Nano Lett.* **2015**, *15*, 3692.

[26] X. Li, F. Cao, D. Yu, J. Chen, Z, Sun, Y. Shen, Y. Zhu, L. Wang, Y. Wei, Y. Wu, H. Zeng, *Small* **2017**, *13*, 1603996.

[27] S. Chen, K. Roh, J. Lee, W. K. Chong, Y. Lu, N. Mathews, T. C. Sum, A. Nurmikko, *ACS Nano* **2016**, *10*, 3959.

[28] C. Huang, W. Sun, S. Liu, S. Li, S. Wang, Y. Wang, N. Zhang, H. Fu, S. Xiao, Q. Song, *Laser & Photonics Rev.* **2019**, *13*, 1800189.

[29] T. K. Fryett, Y. Chen, J. Whitehead, Z. M. Peycke, X. Xu, A. Majumdar, *ACS Photonics* **2018**, *5*, 2176.

[30] Y. Chen, A. Ryou, M. R. Friedfeld, T. Fryett, J. Whitehead, B. M. Cossairt, A. Majumdar, *Nano Lett.* **2018**, *18*, 6404.

[31] M. Khan, T. Babinec, M. W. McCutcheon, P. Deotare, M. Lončar, *Opt. Lett.* **2011**, *36*, 421.

[32] Y. Li, J. Zhang, D. Huang, H. Sun, F. Fan, J. Feng, Z. Wang, C. Z. Ning, *Nat. Nanotechnol.* **2017**, *12*, 987.

[33] Y. Chen, J. Whitehead, A. Ryou, J. Zheng, P. Xu, T. Fryett, A. Majumdar, *Opt. Lett.* **2019**, *44*, 3058.

[34] Y. Wang, X. Li, J. Song, L. Xiao, H. Zeng, H. Sun, *Adv. Mater.* **2015**, *27*, 7101.

[35] Y. Wang, X. Li, V. Nalla, H. Zeng, H. Sun, *Adv. Funct. Mater.* **2017**, *27*, 1605088.

[36] X. He, P. Liu, H. Zhang, Q. Liao, J. Yao, H. Fu, *Adv. Mater.* **2017**, *29*, 1604510.

[37] Z. Hu, Z. Liu, Y. Bian, D. Liu, X. Tang, W. Hu, Z. Zang, M. Zhou, L. Sun, J. Tang, Y. Li, J. Du, Y. Leng, *Adv. Opt. Mater.* **2017**, *5*, 1700419.

[38] C. Huang, C. Zou, C. Mao, K. L. Corp, Y. Yao, Y.J. Lee, C. W. Schlenker, A. K. Y. Jen, L. Lin, *Acs Photonics* **2017**, *4*, 2281.

[39] R. Su, C. Diederichs, J. Wang, T. C. H. Liew, J. Zhao, S. Liu, W. Xu, Z. Chen, Q. Xiong, *Nano Lett.* **2017**, *17*, 3982.

[40] Z. Liu, J. Yang, J. Du, Z. Hu, T. Shi, Z. Zhang, Y. Liu, X. Tang, Y. Leng, R. Li, *ACS Nano* **2018**, *12*, 5923.

[41] Z. Wang, Y. Ren, Y. Wang, Z. Gu, X. Li, H. Sun, *Appl. Phys. Lett.* **2019**, *115*, 111103.

[42] J. Gong, Y. Wang, S. Liu, P. Zeng, X. Yang, R. Liang, Q. Ou, X. Wu, S.Zhang,

*Opt. Express* **2017**, *25*, A1154.

[43] Y. Fu, H. Zhu, C. C. Stoumpos, Q, Ding, J. Wang, M. G. Kanatzidis, X. Zhu, S. Jin, *ACS Nano* **2016**, *10*, 7963.

[44] X. Wang, M. Shoaib, X. Wang, X. Zhang, M. He, Z. Luo, W. Zheng, H. Li, T. Yang, X. Zhu, L. Ma, A. Pan, *ACS Nano* **2018**, *12*, 6170.

[45] B. Zhou, M. Jiang, H. Dong, W. Zheng, Y. Huang, J. Han, A. Pan, L. Zhang, *ACS Photonics* **2019**, *6*, 793.

[46] Q. Liao, X. Jin, H. Zhang, Z. Xu, J. Yao, H. Fu, *Chem. Int. Ed. Engl.* **2015**, *54*, 7037.

[47] S. K. Cohen, S. R. Forrest, *Nat. Photonics* **2010**, *4*, 371.

[48] H. Utzat, W. Sun, A. E. K. Kaplan, F. Krieg, M. Ginterseder, B. Spokoyny, N. D. Klein, K. E. Shulenberger, C. F. Perkinson, M. V. Kovalenko, M. G. Bawendi, Science **2019**, *363*, 1068.

[49] M. Bayer, *Annalen der Physik* **2019**, *531*, 1900039.